\documentclass[12pt]{article}
\usepackage[dvips]{graphicx}
\usepackage{amsmath,amssymb}
\usepackage{bm}

\title{BSD1}
\author{}

\begin{document}
{\pagestyle{empty}
\rightline{September 2007}
\rightline{~~~~~~~~~}
\vskip 1cm
\centerline{\large \bf Regularizations of the Euler product representation for zeta functions}
\centerline{\large \bf and the Birch--Swinnerton-Dyer conjecture}
\vskip 1cm
\centerline{{Minoru Fujimoto\footnote{E-mail address: 
             cayce@eos.ocn.ne.jp}} and
            {Kunihiko Uehara\footnote{E-mail address: 
             uehara@tezukayama-u.ac.jp}}}
\vskip 1cm
\centerline{\it ${}^1$Seika Science Research Laboratory,
Seika-cho, Kyoto 619-0237, Japan}
\centerline{\it ${}^2$Department of Physics, Tezukayama University,
Nara 631-8501, Japan}
\vskip 2cm

\centerline{\bf Abstract}
\vskip 0.2in

  We consider a variant expression to regularize the Euler product representation of 
the zeta functions, where we mainly apply to that of the Riemann zeta function 
in this paper. 
The regularization itself is identical to that of the zeta function of 
the summation expression, but the non-use of the M\"oebius function enable us 
to confirm a finite behavior of residual terms which means an absence of 
zeros except for the critical line. 
Same technique can be applied to the $L$-function associated to 
the elliptic curve, and we can deal with the Taylor expansion at the pole 
in critical strip which is deeply related to the Birch--Swinnerton-Dyer conjecture. 

\vskip 0.4cm\noindent
PACS number(s): 02.30.-f, 02.30.Gp, 02.30.Lt

\hfil
\vfill
\newpage}
\setcounter{equation}{0}
\addtocounter{section}{0}
\section{Introduction}
\hspace{\parindent}

  In the situation that the regularization by way of the Riemann zeta function 
have been successful with some physical applications, 
we proposed a regularization technique\cite{Fujimoto1} and applied that to 
the Euler product of zeta functions.
The definition of the Riemann zeta function is 
\begin{equation}
  \zeta(z)=\sum_{n=1}^\infty\frac{1}{n^z}
          =\prod_{n=1}^\infty\left(1-\frac{1}{{p_n}^z}\right)^{-1}
\label{e101}
\end{equation}
for $\Re z>1$, where the right hand side is the Euler product representation 
and $p_n$ is the $n$-th prime number. 

  In another way the Euler product representation was used for getting formulae 
related with prime number products, for example, $\displaystyle{\prod_n^\infty p_n=4\pi^2}$
\cite{Guy}\cite{Garcia}. 
This formula does, of course, make sense under a regularization procedure and 
the regularized zeta function would be utilized at that time. 
In these streams, an exponential expression of a part of the Euler product can be 
described by using M\"oebius function in the definition of 
the Artin-Hasse exponential\cite{Koblitz} such as
\begin{equation}
  e^{P(z)}=\prod_{n=1}^\infty \zeta(nz)^\frac{\mu(n)}{n},
\label{e102}
\end{equation}
where $\displaystyle{P(z)\equiv\sum_{n=1}^\infty\frac{1}{{p_n}^z}}$ and 
$\mu(n)$ is M\"oebius function.

  The regularized zeta function is well defined even in the critical strip $0<\Re z<1$,  
we adopt a notation such as $\hat{\zeta}(z)$ for the regularized functions 
not to confuse with ones for $\Re z>1$. 
For example, the zeta function by the definition using an alternating summation,
\begin{equation}
  \hat\zeta(z)=\frac{1}{1-2^{1-z}}\sum_{n=1}^\infty\frac{(-1)^{n-1}}{n^z}
\label{e103}
\end{equation}
is well regularized in the critical strip, where we use $\hat{\zeta}(z)$ notation.
Hereafter we are only interested in the region $\Re z\ge\frac{1}{2}$ 
for the Riemann zeta function, because the functional equation ensures us that 
the regularized nature of the zeta function for the other half plane $\Re z<\frac{1}{2}$. 

  In order to get the regularized expression of the Euler product for the Riemann zeta function, 
we consider a variant expression not using M\"oebius function 
but the prime numbers in section 2. 
We also deal with the method of the dipole cancellation limit developed in the previous work\cite{Fujimoto1}
for the same expression in section 2.
Same technique for regularized expression can be applied to the Euler product 
for the $L$-function associated to the elliptic curve, 
and we can deal with the Taylor expansion at the pole in critical strip 
which is deeply related to the Birch--Swinnerton-Dyer (BSD) conjecture. 
So in \S 3 we discuss the BSD conjecture and concluding remarks.

\hskip 5mm
\section{Regularization for the Euler product representation}
\hspace{\parindent}

As is stated in the previous section, the Riemann zeta function is expressed 
in the Euler product representation for $\Re z>1$, 
\begin{equation}
  \zeta(z)=\prod_{n=1}^\infty\left(1-\frac{1}{{p_n}^z}\right)^{-1}.
\label{e201}
\end{equation}
  We can transform the equation to followings by taking logarithms of both hand sides:
\begin{eqnarray}
\log\zeta(z)&=&\sum_{n=1}^\infty\left(
					\frac{1}{{p_n}^z}+\frac{1}{2{p_n}^{2z}}+
					\frac{1}{3{p_n}^{3z}}+\frac{1}{4{p_n}^{4z}}+
					\frac{1}{5{p_n}^{5z}}+\cdots
			\right)
\label{e202}\\
		&=&	P(z)+\frac{1}{2}P(2z)+
			\frac{1}{3}P(3z)+\frac{1}{4}P(4z)+
			\frac{1}{5}P(5z)+\cdots
\label{e203}\\
		&=&	P(z)+R(z)
\label{e204}
		,
\end{eqnarray}
where we have used $\displaystyle{P(z)=\sum_{n=1}^\infty\frac{1}{{p_n}^z}}$ in the second line. 
When we think about functions in the region $\Re z>\frac{1}{2}$, 
we easily recognize that each term after the second converges in Eq.(\ref{e203}). 
Moreover, it can be shown that the infinite sums after the forth term converges 
by ceiling the value of the formula $\displaystyle{\sum_{k=2}^\infty\frac{\zeta(k)-1}{k}=1-\gamma}$ with the Euler constant $\gamma$.
Thus we express only the first term $P(z)$ as the term for a regularization using 
the zeta function and prime numbers. 
The expression for $P(z)$ for $\Re z>1$ can be got by adding or subtracting a term of 
$\displaystyle{\sum\frac{1}{\prod p}\log\zeta(\prod p\ z)}$, 
\begin{eqnarray}
  P(z)&=&\log\zeta(z)
        -\sum_{i=1}^\infty\frac{1}{p_i}\log\zeta(p_iz)
        +\sum_{1\le i<j}^\infty\frac{1}{p_ip_j}\log\zeta(p_ip_jz)\nonumber\\
      &&-\sum_{1\le i<j<k}^\infty\frac{1}{p_ip_jp_k}\log\zeta(p_ip_jp_kz)+-\cdots,
\label{e205}
\end{eqnarray}
where a coefficient of $\displaystyle{\frac{1}{m}P(mz)}$ term in Eq.(\ref{e205}) for 
$m={p_{i_1}}^{\alpha_1}{p_{i_2}}^{\alpha_2}\cdots{p_{i_n}}^{\alpha_n}$ 
cancels out by using the relation $\displaystyle{\sum_{i=1}^n(-1)^n{}_nC_i=0}$. 
A regularization of $P(z)$ in the region of $\frac{1}{2}<\Re z<1$ is performed 
at the same time that the analytic continuation of $\zeta(z)$ to the region 
$\Re z<1$ to hold the equation (\ref{e205}) above. 
In order to see the situation, we introduce 
\begin{eqnarray}
  \hat{\zeta}(z)=\lim_{n\to\infty}\hat{\zeta}_n(z)
                &=&\lim_{n\to\infty}\left\{\zeta_n(z)-\frac{n^{1-z}}{1-z}\right\},\\
\label{e208}
  \hat{P}(z)=\lim_{n\to\infty}\hat{P}_n(z)
            &=&\lim_{n\to\infty}\left\{P_n(z)-\int_1^n\frac{dt}{\{\pi^{-1}(t)\}^z}\right\}+O(1)\\
\label{e209}
			&=&\lim_{n\to\infty}\left\{P_n(z)-\int_1^n\frac{dt}{(t\log t)^z}\right\}+O(1),
\label{e210}
\end{eqnarray}
where $\displaystyle{\zeta_n(z)=\sum_{k=1}^n\frac{1}{k^z}}$ and 
$\displaystyle{P_n(z)=\sum_{k=1}^n\frac{1}{{p_k}^z}}$, 
and we define $R_n(z)\equiv\log\zeta_n(z)-P_n(z)$, 
which is finite even in the limit of $n\to\infty$.

We will show the finiteness of $\hat{P}(z)$ for $\frac{1}{2}<\Re z<1$
\begin{eqnarray}
  P_n(z)&=&\log\left\{\frac{n^{1-z}}{1-z}+\hat{\zeta}_n(z)\right\}-R_n(z)\\
        &=&\log\frac{n^{1-z}}{1-z}+\log\left\{1+\frac{(1-z)\hat{\zeta}_n(z)}{n^{1-z}}\right\}-R_n(z).
\label{e211}
\end{eqnarray}
In the last line, as 
$\displaystyle{\lim_{n\to\infty}\log\left\{1+\frac{(1-z)\hat{\zeta}_n(z)}{n^{1-z}}\right\}=0}$ 
under the condition $z\ne1, \zeta(z)\ll\infty$, we can conclude 
\begin{eqnarray}
  P_n(z)&=&\log\frac{n^{1-z}}{1-z}\\
        &=&(1-z)\log n+\log\frac{1}{1-z},\\
  \hat{P}(z)&=&\log\frac{1}{1-z}+O(1).
\label{e212}
\end{eqnarray}

  After all for $\frac{1}{2}<\Re z<1$, $\hat{P}(z)$ is finite, so 
$\hat{\zeta}(z)=-e^{\{\hat{P}(z)+R(z)\}}$ cannot vanish in this strip, 
which means that the regularized Riemann zeta function does not 
take zeros for the critical strip. 
The negative sign yields from the fact that the zeta function 
$\hat{\zeta}(z)$ has the pole of order one at $z=1$, 
the relation $\hat{\zeta}(z)\sim e^{\hat{P}(z)}$ changes a sign for $\Re z<1$.

\vskip 5mm

  From now on we regularize $P(z)$ by 
the method of the dipole cancellation limit. 
The dipole equation for $P_n(z)$ is 
\begin{equation}
  \{1-\alpha_k(z)\}P_k(z)+\alpha_{k+1}(z)P_{k+1}(z)=0,
\label{e301}
\end{equation}
where $\alpha_k(z)$ is a ratio of the internal division for the value of 
$k$-th term $P_k(z)$. 
We can find the solution given by
\begin{equation}
  \alpha_1(z)=P_1(z)^{-1}\left\{\sum_{k=1}^{n-1}P_k(z)+\alpha_n(z)P_n(z)\right\},
\label{e302}
\end{equation}
and the expression $\alpha_k(z)=-P_k(z)p_{k+1}^z$ is satisfied with Eq.(\ref{e301}). 
Thus the regularized $\hat{P}(z)$ is
\begin{equation}
  \hat{P}(z)=\lim_{n\to\infty}2^z\left\{\sum_{k=1}^{n-1}P_k(z)-(P_n(z))^2p_{n+1}^z\right\},
\label{e303}
\end{equation}
where as you will see that $\hat{P}(z)$ is same as one in Eq.(\ref{e212}). 
As the regularized value by the method of the dipole cancellation limit 
is finite, this $\hat{P}(z)$ gives the expression of the limit of $n\to\infty$ 
for 
$$
  e^{\{P_n(z)+R_n(z)\}}=\zeta_n(z),
$$
and the Riemann hypothesis is satisfied again. 
  As we have seen in this section, the method of the dipole cancellation limit 
can be applicable even to the complicated function and remove the leading 
divergence easily.


\vskip 5mm
\section{Discussion and conclusion}
\hspace{\parindent}
  For an elliptic curve, we can estimate $N(A)$ the number of rational points 
on the curve smaller than the height $A$. 
Thus, the fact $E(\bm{Q})\simeq \bm{Z}^r\oplus T$ 
for a finite group leads to
\begin{equation}
  N(A)\sim C(\log A)^{r/2}\ \ \ \text{ in the limit of }\  A\to\infty,
\label{e401}
\end{equation}
where $C$ is some positive constant.
  The Birch--Swinnerton-Dyer conjecture\cite{Birch1}\cite{Birch2} claimes 
that the existence of an asymptotic expression
\begin{equation}
  \prod_{p<x}\frac{N_p}{p}\sim C'(\log x)^r\ \ \ \text{ in the limit of }\  x\to\infty,
\label{e402}
\end{equation}
where $C'$ is a constant related to $C$, and that the $L$-function $L(E,z)$ has 
the order $r$-th zero at $z=1$ even when $E(\bm{Q})$ is the infinite group. 
The BSD conjecture demands that $L(E,z)$ should be continued into the left side 
beyond $\Re z=\frac{3}{2}$.
\begin{equation}
  L(E,z)\sim\prod_{p\nmid\Delta}\left(1-\frac{a_p}{p^z}+\frac{p}{p^{2z}}\right)^{-1},
\label{e403}
\end{equation}
where $\Delta$ is the discriminant of the elliptic curve and 
$a_p$ is related to $p$ and the elliptic curve.

  The regularization stated in the previous section, namely, the regularization of 
the Euler product representation for the Riemann zeta function 
can be applicable to the $L$-functions or 
generalized Euler product. 
  Specially when we apply it to the $L$-function associated to 
the elliptic curve, (in which the Riemann hypothesis holds\cite{Fujimoto2},)
the BSD conjecture can be taken into account.
  When we take the latest result by Gross and Zagier\cite{Gross} {\it et al.}
for the BSD conjecture
\begin{equation}
  L(E,1)\ne0\ \ \text{ or }\ \ L'(E,1)\ne0,
\label{e403}
\end{equation}
$L(E,1)\ne0$ follows by the regularization of the Euler product and
$L'(E,1)\ne0$ is satisfied by the generalized Riemann hypothesis 
even if $L(E,1)=0$, which means the BSD conjecture is demonstrated.

\vskip 5mm

\end{document}